\documentclass[journal]{IEEEtran}
\usepackage{cite}
\ifCLASSINFOpdf
\else
\fi
\usepackage{array}
\usepackage{mdwmath}
\usepackage{mdwtab}
\usepackage{graphicx}
\usepackage{euscript}
\usepackage{amsfonts}
\usepackage{bigstrut}
\usepackage{tabularx}
\usepackage{cite}
\usepackage{amssymb}
\usepackage{nomencl}
\usepackage{hyperref}
\usepackage{algorithm}
\usepackage[noend]{algpseudocode}
\makeatletter
\def\BState{\State\hskip-\ALG@thistlm}
\makeatother
\usepackage{enumitem}
\usepackage{upgreek}
\usepackage{dsfont}
\usepackage{amsthm}
\usepackage{amssymb}
\usepackage{xcolor}
\usepackage{mathtools}

\usepackage{hyperref}
\usepackage{epstopdf} 
\usepackage{bbm}
\DeclareMathOperator*{\argmax}{argmax} 
\usepackage{microtype}
\usepackage{booktabs}

\usepackage{multirow}
\usepackage{textcomp, gensymb}

\DeclareMathOperator{\asin}{asin}

\DeclareMathOperator{\atantwo}{atan2}

\title{Spiking Neural Networks for Detecting Satellite-Based Internet-of-Things Signal}

\author{Kosta Dakic$^\star$,~\IEEEmembership{Graduate~Student Member,~IEEE}, Bassel~Al~Homssi,~\IEEEmembership{Member,~IEEE},\\ Sumeet Walia,~\IEEEmembership{Member,~IEEE}, and Akram Al-Hourani$^\star$,~\IEEEmembership{Senior~Member,~IEEE},
\thanks{$^\star$Corresponding authors: K.~Dakic and A. Al-Hourani, with the School of Engineering, RMIT University, Melbourne, VIC 3000. E-mail: kosta.dakic@ieee.org, akram.hourani@rmit.edu.au.}
 }

\begin{document}

\maketitle

\begin{abstract}
With the rapid growth of IoT networks, ubiquitous coverage is becoming increasingly necessary. Low Earth Orbit (LEO) satellite constellations for IoT have been proposed to provide coverage to regions where terrestrial systems cannot. However, LEO constellations for uplink communications are severely limited by the high density of user devices, which causes a high level of co-channel interference. This research presents a novel framework that utilizes spiking neural networks (SNNs) to detect IoT signals in the presence of uplink interference. The key advantage of SNNs is the extremely low power consumption relative to traditional deep learning (DL) networks. The performance of the spiking-based neural network detectors is compared against state-of-the-art DL networks and the conventional matched filter detector. Results indicate that both DL and SNN-based receivers surpass the matched filter detector in interference-heavy scenarios, owing to their capacity to effectively distinguish target signals amidst co-channel interference. Moreover, our work highlights the ultra-low power consumption of SNNs compared to other DL methods for signal detection. The strong detection performance and low power consumption of SNNs make them particularly suitable for onboard signal detection in IoT LEO satellites, especially in high interference conditions.
\end{abstract}

\section{Introduction}~\label{Section_I}

IoT use cases are growing at an unprecedented rate, where terrestrial networks are unable to supply coverage for applications like smart farming and parcel tracking in rural, remote areas, and areas with extreme environments~\cite{RN651}. To accommodate the growing need for global coverage, low earth orbit (LEO) satellite constellations have been of great interest to the research community and industry alike~\cite{RN652,RN646,RN635} where many large companies such as OneWeb, SpaceX, and Amazon~\cite{oneWeb,starlink_2022,kupier} are currently deploying mega satellite constellations. For an alternative method to provide global coverage, geosynchronous orbit (GEO) satellites could also be used. However, LEO satellite links have a lower overall propagation delay and a lower propagation loss relative to GEO links~\cite{RN635}. Moreover, GEO satellites require a higher cost for deployment, and a typical GEO link requires large transmit/receive antennas and stronger transmit power, which makes them inadequate for networks constrained with low cost and low energy such as IoT networks.

However, LEO satellite networks suffer from high Doppler due to the satellite motion relative to the ground user as well as a high level of co-channel interference. The high level of interference is caused by other devices operating in the same time-frequency resources due to the wide coverage area of the satellite. High interference in multiple access LEO networks is caused by a large number of co-channel transmissions, particularly in shared bands such as industrial, scientific, and medical (ISM) band~\cite{8403749} which is open to the general public for transmission and does not require paid licensing. The devices in these bands lack standardization and coordination which exacerbates interference~\cite{RN652}. In these interference-limited scenarios, typical detection (demodulation) methods result in a high symbol error rate (SER). To mitigate co-channel interference methods such as frequency reuse, dynamic spectrum allocation, and beam multiplexing techniques are typically used. Nevertheless, the interference levels are still large, to address the high SER brought about by co-channel interference, deep learning (DL) signal detection has been proposed in~\cite{RN648,RN324,RN738} to efficiently detect signals in interference-limited scenarios.

Generic DL methods such as the simple artificial neural network (ANN) and the convolutional neural network (CNN) have shown strong performance in various applications ranging from natural language processing (NLP)~\cite{RN708} to image classification~\cite{3065386}. However, they consume a substantial amount of power depending on the size of the architecture~\cite{RN720}. Due to the large power consumption, generic DL networks are typically designed to perform localized computing tasks, i.e. edge computing, and thus do not suit resource-limited devices. On the other hand, the human brain is able to process complicated dynamic events at only an average of 20 W power~\cite{RN675}. On the other hand, traditional processing such as the largest neural network (NN) model, dubbed the Megatron-Turing natural language generation (MT-NLG) model with 530 billion parameters~\cite{smith2022using} uses multiple high-end GPUs which consume many orders of magnitude more power than the brain. Furthermore, a NN with 530 billion parameters is much less complex than the human brain which has around 100 billion neurons and 100 trillion synapses. To emulate the behavior of a single biological neuron, a 5-8 layer NN with hundreds of neurons per layer is required according to the authors in~\cite{RN721}. Consequently, the third generation of NN models takes inspiration from the spiking behavior in human brains. Accordingly, the Spiking neural network (SNN) is set to dramatically reduce power consumption~\cite{RN707} by encoding data with intermittent spiking that only produces an output when the input exceeds a certain threshold. This means that SNNs only consume power when there is an event (i.e. a spike) at the neuron, thus raising the level of efficiency closer to that of the biological brains. Due to the exceptional energy efficiency of the SNN relative to conventional DL networks, an SNN-based IoT signal detector would be ideal for a satellite receiver due to its limited power resources.

In this paper, we demonstrate the performance of spiking models and their applicability to modern communications systems, specifically for chirp signal detection. We present a comparison of traditional detection, conventional DL models, and spiking models for satellite uplink signal detection of devices that utilize the chirp-based modulation scheme. As a practical implementation, we consider the LoRa modulation scheme which is a popular chirp-based modulation and a technology of interest in terrestrial IoT networks. Nevertheless, the proposed detector can be adapted to many other signal types. The comparison evaluates the performance of how conventional and spiking models can improve the SER in an environment with high levels of co-channel interference. Additionally, the performance assessment relies on the emulation of a LEO satellite constellation with ground users that employ a chirp signal modulation scheme, where emulation of the chirp signal is performed with a MATLAB-scripted emulator~\cite{9395074,Emulator} developed earlier by our research team. The results show a considerable improvement in the SER of conventional DL and spiking-based signal detectors compared to traditional non-coherent detection in the uplink of a LEO constellation. The contribution of this work is summarized as follows,
\begin{itemize}
    \item It demonstrates the applicability of power-efficient spiking-based networks in detecting uplink IoT signals in a LEO satellite scenario with high co-channel interference levels.
    \item It assesses the power efficiency performance of spiking networks based on neuromorphic hardware relative to conventional DL models running on standard hardware.
    \item It develops a synchronization scheme that is novel in the IoT-over-space context.
    \item It tests the idea from~\cite{RN648} which shows a hybrid network that switches between the conventional matched filter and DL signal detection based on the inferred interference level to further increase the energy efficiency of the spiking-based receivers by hybridizing the network with conventional detection techniques.
\end{itemize}

The rest of this paper is organized as follows; In Section~\ref{Section_II} we give some related works. In Section~\ref{Section_III} we describe the SNN. In Section~\ref{Section_IV} we detail the system model for emulating a LEO constellation for uplink transmission. We describe the geometric model, which covers the satellite constellation, satellite beam footprint, and the distribution of the user devices. Furthermore, we discuss the wireless access model and the channel.  Section~\ref{Section_V} describes the spiking-based detection networks used in this paper. In Section~\ref{Section_VI} we cover the physical layer of chirp-based modulation (LoRa). We introduce our proposed synchronization method in Section~\ref{Section_VII}. We show and discuss the results in Section~\ref{Section_VIII}. Finally, the paper is concluded in Section~\ref{Section_IX}.

\section{Related Works}~\label{Section_II}
The uplink performance of LEO IoT networks has been analyzed using stochastic geometry in~\cite{RN644} and simulation in~\cite{rs12101666}. In these research works co-channel interference is concluded to heavily limit the performance. To combat the performance degradation arising from co-channel interference, many interference mitigation techniques have been studied in past research works. One method is to use coordinated multiple-access techniques, i.e. scheduling similar to cellular networks~\cite{RN739}. However, using coordinated multiple-access techniques is difficult to implement in LEO satellite constellations given the high Doppler shift as well as the higher propagation delay compared to conventional terrestrial networks. Other interference mitigation approaches such as Cognitive Radio (CR)~\cite{RN647} have been utilized, however, CR is not ideal for low-power user devices due to the extra power needed for spectrum sensing.

One prominent technology that is positioned for IoT-over-satellite is chirp-base modulation. The performance of LoRa (as the implementation of chirp-based modulation) under interference conditions has been investigated in multiple studies, such as in~\cite{8903531,RN644,RN308,unknown}. While traditional detection of LoRa shows some robustness to interference,  the network cannot be directly scaled to practical satellite scenarios given typical traditional non-coherent detector receivers. An uplink LEO satellite communication scenario is an example of a high interference environment, where the satellite receiver has to deal with a high volume of radio frames, due to its inherent large swath of coverage. An IoT-over-satellite scenario concentrating on LoRa modulation is explored in~\cite{RN685}, where it is concluded that high volumes of LoRa traffic significantly deteriorate the system’s performance. In the study referenced as~\cite{RN682}, the authors propose a folded chirp shift keying (FCSK) system that, like LoRa, employs chirp-based modulation for transmitting low-bit-rate data. Although their approach demonstrates enhanced resilience to Doppler effects, FCSK's detection error rate is inferior to that of LoRa. This is primarily attributed to FCSK's variable chirp rate, which renders it more vulnerable to interference from other chirp rates. Nonetheless, receivers need to be significantly redesigned such that they could deal with non-Gaussian impairments. One promising direction is the use of DL tools, as shown in~\cite{RN648}, which leverage DLs robustness against non-Gaussian impairments to enhance signal receivers in an interference-limited regime. 

DL has been shown to be a great tool for dealing with stochastic impairments for signal detection in wireless communication~\cite{RN648,RN324,RN738}. DL has also been utilized to achieve a performance gain by optimizing an interference mitigation algorithm~\cite{RN555}. However, DL as a tool for interference mitigation has not been extensively showcased in a satellite-based Internet of Things (S-IoT) scenario. For a satellite receiver, a DL receiver has been shown in~\cite{RN650}, however, the work focuses on high throughput transmissions and the non-linear impairments produced by the hardware. While conventional DL works well at detecting signals with non-Gaussian noise, the high power requirements of complex DL models are not ideal for use in systems with limited power availability such as onboard processors in LEO satellites. A spiking-based system would reduce the energy requirements while still providing high-performance characteristics for signal detection~\cite{RN707}.

\section{Spiking Neural Networks}~\label{Section_III}
An advantage that SNNs have over generic NNs is their addition of a spatiotemporal dimension in the spike trains instead of single element vectors used in NN that carry the weights and biases. The spike trains allow the SNN to be effective at carrying a lot of information, by using fewer neurons relative to generic NNs~\cite{RN660}. As a result, an SNN would perform particularly strongly at processing real-time, continuous, and temporally rich data streams, such as wireless signals~\cite{satmlmag}. Also, the SNN network is only energized when the spikes are generated/received, unlike typical NNs where the entire network needs to be continuously run for every new input.

These advantages have been used for several previous research works such as in~\cite{RN679}, where an unsupervised SNN was used in automatic modulation classification (AMC). Another related example can be found in~\cite{RN680}, where the authors efficiently perform AMC on RF domain data using an SNN and improve memory utilization by more than three orders of magnitude. Research work on using SNNs for signal detection is proposed in~\cite{RN681} to improve human body communication by harnessing the aforementioned low power consumption and superior performance on spatiotemporal data of the SNN. Another research work for detecting signals with SNNs is demonstrated in~\cite{RN711}, where an SNN is used for detecting radar signals. Notwithstanding, to the best of our knowledge, there has not been any research work showing the prospect of SNNs for signal detection of uplink for IoT-over-satellite communications. The prime benefit of an SNN-based receiver would be its ability to deal with non-Gaussian interference while simultaneously exhibiting low power consumption.

\begin{figure}[!t]
{\centering
\includegraphics[width=\linewidth]{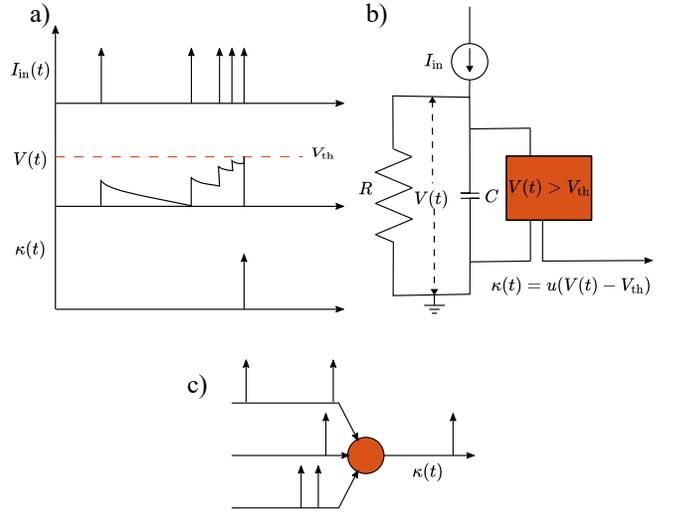}
\caption{An illustration of the leaky integrate-and-fire neuron model. }
\label{Fig_LIF}}
\footnotesize
\end{figure}

The spiking neuron is heavily inspired by biologically plausible neuron models such as the Hodgkin-Huxley model~\cite{RN709}. However, while the Hodgkin-Huxley model is accurate, it is difficult to compute. Consequently, leaky integrate-and-fire (LIF) neuron models can be used instead. The LIF neuron model takes the sum of weighted input pulses, which in turn are integrated over time with a certain exponential decay, i.e., \emph{leakage}. An illustration is shown in Fig.~\ref{Fig_LIF}(a) showing that a membrane potential is accumulated from the input current $I_\mathrm{in}$ and if the integrated value exceeds a chosen threshold $V_\mathrm{th}$, then the LIF neuron will fire a voltage spike $\kappa(t)$. The relationship between the output $\kappa$ and the voltage $V(t)$ can be represented as follows,
\begin{equation}
    \kappa(t) = u\left(V(t)-V_\mathrm{th}\right) , \label{LIF_Eq}
\end{equation}
where $u(.)$ is the Heaviside step function. When a spike is triggered, the membrane potential should be reset. Next, the illustration in Fig.~\ref{Fig_LIF}(b) shows the equivalent Lapique RC circuit, which is composed of the membrane capacitance C and the membrane resistance R. The circuit can be mathematically modeled as follows,
\begin{equation}
    RC\frac{dV(t)}{dt} = -V(t) + I_\mathrm{in}(t)R ,
\end{equation}
where the solution to the differential equation is~\cite{neuronal},
\begin{equation}
    V(t) = \frac{R}{RC} \int_{0}^{\infty} \text{exp}\left(-\frac{q}{RC}\right)I(t-q)~\text{d}q .
\end{equation}
Finally, Fig.~\ref{Fig_LIF}(c) shows how each connection to the neuron body accumulates outputs, then $\kappa(t)$ generates a voltage spike. Ultimately, the temporal information of the spiking data carries the learned attribute from the input to the output of the spiking neural network. The temporal information of the spiking data evolves over time as the network learns.

\begin{figure}[!t]
	{\centering
		\includegraphics[width=\linewidth]{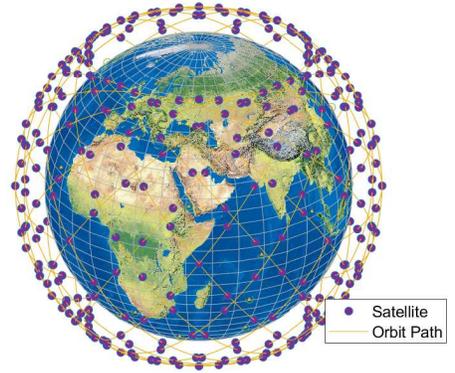}
		\caption{ Walker-Delta orbit simulation using 384 satellites with an orbital inclination of $53^\circ$. }
		\label{Fig_orbit}}
	\footnotesize
\end{figure}

\section{System Model}~\label{Section_IV}
The simulation model developed in this work emulates randomly distributed ground IoT devices that operate using the chirp modulation scheme. Each ground IoT device is configured to connect to the nearest satellite if the device lies within the satellite’s footprint, where each satellite is part of a large constellation. The satellites are simulated to orbit a perfectly spherical Earth as an approximation. The model assumes all IoT devices use chirp-based modulation without loss of generality, which is performed using a Matlab scripted emulator~\cite{Emulator} to generate the chirp-based signals. Note that for this work we utilize the LoRa physical layer, however, the framework can support other physical layer technologies, and LoRa is used as an example. The following sections further describe the geometric model and the channel model.
\subsection{Geometric Model}
\begin{figure}[!t]
{\centering
\includegraphics[width=\linewidth]{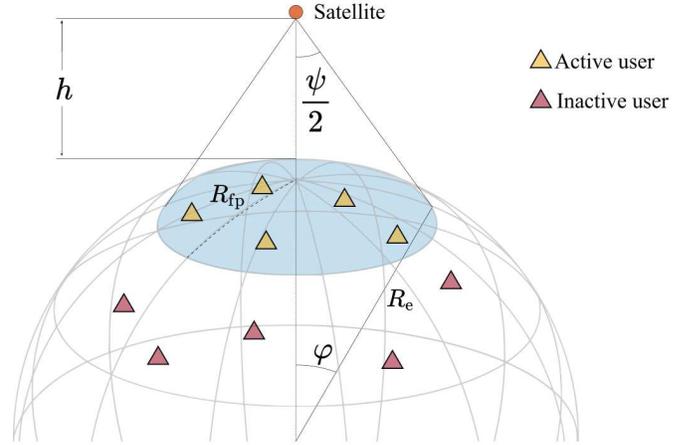}
\caption{ LEO satellite scenario showing the concept of Earth-centered zenith angle $\varphi$ and the satellite beamwidth $\psi$. }
\label{Fig_Geometry}}
\footnotesize
\end{figure}

We consider a practical satellite constellation for this work, where the Walker-Delta constellation is utilized. The Walker-Delta constellation is used to demonstrate practical network performance as the constellation is adopted by large satellite constellation companies, such as Starlink~\cite{starlink_2022}. The Walker-Delta constellation is described by parameters ${h},{i},{N},{P}~\mathrm{and},~{F}$, where $h$ is the altitude of the satellite, $i$ us the satellite orbital inclination angle, $N$ is the total number of satellites, $P$ is the number of orbital planes and, $F$ is the phasing parameters used to describe the phase difference between satellites in consecutive orbital planes. Each orbital plane has an equal number of satellites $N/P$ where the satellites are uniformly spaced. A snapshot of an example of a Walker-Delta constellation in the simulation is shown in Fig.~\ref{Fig_orbit}. For the user to be able to communicate with the satellite, it needs to lay within the satellite footprint. For a simplified conical antenna pattern, The footprint is governed by the satellite effective beamwidth~\cite{9684552}, denoted as $\psi$, which dictates the footprint projection on the Earth's surface. The footprint projection is assumed to be an ideal spherical cap bounded by an earth-centered zenith angle, denoted as $\varphi$ (as indicated in Fig.~\ref{Fig_Geometry}). Using simple geometric reasoning, the area of the spherical cap of the beam is calculated as follows,
\begin{equation}
	A_\mathrm{fp} = 2\pi R_\mathrm{e}^2\left(1-\cos\varphi\right) , 
\end{equation}
and the earth-centered zenith angle is calculated using the law of sines as follows~\cite{9422812},
\begin{equation}
    \varphi = \asin\left(\frac{1}{\alpha}\sin\frac{\psi}{2}\right) - \frac{\psi}{2} , 
\end{equation}
where $\alpha = R_\mathrm{e}/R$, $R_\mathrm{e}$ is Earth's average radius, $R = R_\mathrm{e} + h$, and $h$ is the satellite altitude above the Earth's mean sea level. The satellite footprint size is restricted by the horizon. Thus, the maximum effective beamwidth is,
\begin{equation}
    \psi_{\mathrm{max}} = 2\mathrm{asin}\alpha ,
\end{equation}
Moreover, the earth-centered zenith angle when the beamwidth is maximum can be calculated as follows,
\begin{equation}
    \varphi_\mathrm{max} = \mathrm{acos}\left(\frac{R_\mathrm{e}}{R}\right) .
\end{equation}

Then, a spherical cap perimeter is drawn around each satellite footprint to define the boundary, where if a user device is located within the footprint, it is considered active (connected state). The footprint radius of each satellite is calculated as follows,
\begin{equation}
	R_\mathrm{fp} = R_\mathrm{e}{\varphi} .
\end{equation}
For defining the perimeter of the footprint, the latitude and longitude of the footprint boundary need to be calculated with the heading formulae~\cite{RN740} as follows,
\begin{equation}
	\phi_\mathrm{fp} = \asin\left(\sin \phi_\mathrm{sat} \cos \varphi + \cos \phi_\mathrm{sat} \sin \varphi \cos \theta\right) , 
\end{equation}
and the longitude,
\begin{multline}
	\rho_\mathrm{fp} = \rho_\mathrm{sat} + \atantwo(\sin \theta \sin \varphi \cos \phi_\mathrm{sat} , \\	 \cos \varphi - \sin \phi_\mathrm{sat} \sin \phi_\mathrm{fp}) ,
\end{multline}
where $\theta$ is an array from 0 to $2\pi$ with 360 elements and $\phi_\mathrm{sat}$ and $\rho_\mathrm{sat}$ is the latitude and longitude of the satellite's sub-point. A satellite sub-point refers to the point on the Earth's surface directly below a satellite in orbit. An illustration of the geometry of a LEO satellite is shown in Fig.~\ref{Fig_Geometry}.

\begin{figure}[!t]
	{\centering
		\includegraphics[width=\linewidth]{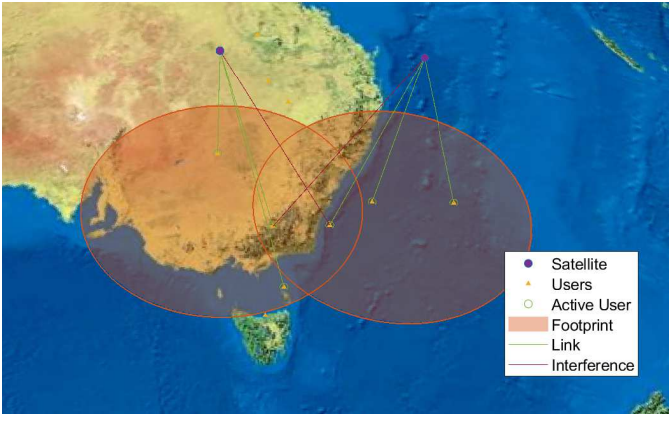}
		\caption{ A simulation snapshot showing the distribution of user devices and the satellite footprint. The red line from the ground user to the satellite indicates that even though the user is not being served by the satellite with the link, it still contributes to the interference. }
		\label{Fig_systemModel}}
	\footnotesize
\end{figure}

For this work, all the user devices are assumed to be homogeneously distributed over the surface of the earth. The devices are distributed using a spherically wrapped Poisson Point Process (PPP), where the density of devices is controlled by $\lambda=D\lambda_\mathrm{o}$. $D$ is the spatial duty cycle of the active devices per second and $\lambda_\mathrm{o}$ is the density of the overall number of devices. Note when we have a large number of users and satellites, a user could be located within more than a single satellite footprint. Consequently, a such user contributes to interference to all these satellites An example of randomly distributed users and links from the active users to the satellite is shown in Fig.~\ref{Fig_systemModel}.

\subsection{IoT Access System}
\begin{figure}[!t]
{\centering
\includegraphics[width=\linewidth]{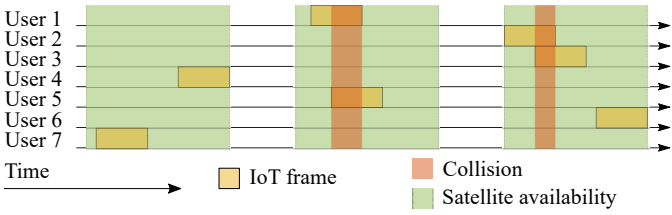}
\caption{ An illustration of the pure-ALOHA access model used for IoT-over-Satellite. }
\label{Fig_accessModel}}
\footnotesize
\end{figure}

For this research work, we are interested in the performance of signal detection in a satellite scenario where received signals are prone to co-channel interference from uncoordinated signal transmissions. One popular access system, dubbed pure-ALOHA, does not take into consideration any scheduling or whether another device is transmitting. As such, pure-ALOHA tends to allow for signal collisions that compromise the signal receiver to accurately detect the signal. Additionally, pure-ALOHA does not take into account the availability of a satellite, thus packets can be lost. However, we only care about signal detection in a satellite scenario, so we put more emphasis on detection rather than satellite availability. Nevertheless, the advantage of pure-ALOHA is its simplicity, it requires a low overhead which allows for less power consumption and therefore a longer battery life for IoT devices. Accordingly, we assume a pure-ALOHA access model for this research work, where each device uses the same channel resources and transmits them as soon as the data is available. As an alternative, scheduled ALOHA could be used, where each user device would require a synchronized clock to regulate the transmission times in a periodical or event-triggered manner. Another option is slotted ALOHA, where each station can transmit at any time, but transmissions are divided into time slots. This also helps to reduce collisions and improve network efficiency, but to a lesser extent than scheduled ALOHA. However, both scheduled and slotted ALOHA requires additional overhead for coordinating the transmissions and use additional power as a consequence. An illustration of the pure-ALOHA access model is shown in Fig.~\ref{Fig_accessModel}.

\subsection{Channel Model}
The received time-series signal at the satellite receiver is depicted as,
\begin{equation}\label{channel_Main}
		y(t) = g\{x_i(t)\} + \underbrace{\sum_{j\neq i}g\{x_j(t)\}}_{\mathrm{interference}} + n(t) ,
\end{equation}
where $x_i(t)$ is the time-series signal received from the target user denoted by the subscript $i$, $x_j(t)$ is the time-series signal from an interfering device $j$, $g\{.\}$ represents the channel function which accounts for both the fading and Doppler shift, and $n(t) \sim \mathcal{CN}\left(0,1\right)$ is the complex zero-mean AWGN.

The largest portion of the satellite link is under free-space conditions and only a small portion undergoes excess path loss such as fading due to near-ground clutter. The fading is also contributed by the rain and atmospheric absorption depending on the operating frequency. In addition, LEO satellites orbit Earth at high speeds causing high Doppler shifts. Furthermore, a random phase shift is considered in this research work as perfect phase estimation at the receiver is difficult to achieve in an IoT packet for LEO satellite uplink. This is because LEO satellites have a relatively fast orbital velocity and a low altitude, which results in a rapidly changing Doppler shift and a high signal attenuation due to atmospheric absorption and scattering. Additionally, co-channel interference can make it difficult for the receiver to estimate the phase of the transmitted signal. Hence, the channel is modeled as follows,
\begin{equation}\label{channel_Response}
    g\{x(t)\} = \sqrt{\frac{P_r}{N_\mathrm{o}}}x(t) \exp(j2\pi \nu(t) + j\phi_\mathrm{s}) ,
\end{equation}
where $N_\mathrm{o}$ is the noise power spectral density, $\nu(t)$ is the Doppler frequency shift, and $\phi_\mathrm{s}$ represents the phase shift of the signal. The received power at the satellite receiver, represented as $P_\mathrm{r}(\varphi,\psi)$ and is obtained as follows,
\begin{multline}\label{channel_Power}
    P_\mathrm{r}(\varphi)\mathrm{[dBm]} = \mathsf{EIRP}\mathrm{[dB]} + G\mathrm{[dB]} \\ - \underbrace{l(\varphi)\mathrm{[dB]} - \eta(\varphi) \mathrm{[dB]}}_{\mathrm{Path-loss}},
\end{multline}
where $G$ is the satellite antenna gain, $l(\varphi)$ is the free-space-path-loss and $\eta(\varphi)$ is the excess path-loss. The satellite antenna gain $G$ is obtained by using the ideal antenna gain expression as follows~\cite{RN644},
\begin{equation}
    \begin{split}
        G & = \frac{\mathrm{spherical~area}}{\mathrm{spherical~cap~area}} \\
        & = \frac{4\pi{R_e}^2}{2\pi{R_e}^2\left(1-\mathrm{cos}\frac{\psi}{2}\right)} = \frac{2}{1-\mathrm{cos}\frac{\psi}{2}} .
    \end{split}
\end{equation}
Since typical IoT devices operate near the L-band~\cite{rs12101666} the excess path-loss model is adopted from~\cite{9257490}. This model assumes a two-state channel; the (i) line-of-sight (LoS) and (ii) non-line-of-sight (nLoS) states are assumed. Accordingly, the excess path-loss in decibels is modeled as a mixed normal random variable as follows,
\begin{multline}
    \eta(\varphi)\mathrm{[dB]} \sim p_\mathrm{LoS}(\varphi)\mathcal{N}(\mu_\mathrm{LoS},\sigma_\mathrm{LoS}^2) \\ + p_\mathrm{nLoS}(\varphi)\mathcal{N}(\mu_\mathrm{nLoS},\sigma_\mathrm{nLoS}^2)
\end{multline}
where $p_\mathrm{LoS}(\varphi) = \exp(-\beta \sin\varphi/[\cos\varphi - \alpha])$~\cite{9509510} is the probability that the link is in the LoS state. Note that $\{\beta,\mu_\mathrm{LoS}, \sigma_\mathrm{LoS}, \mu_\mathrm{nLoS}, \sigma_\mathrm{nLoS}\}$ are propagation parameters that depend on the ground device's environment~\cite{9257490}. On the other hand, the free space path loss only depends on the operating frequency and the distance between the ground device and receiving satellite and is given as follows,
\begin{equation}
    l(\varphi) = \left[{\frac{4\pi d(\varphi) f_\mathrm{c}}{c}}\right]^2,
\end{equation}
where $c$ is the speed of light and $f_\mathrm{c}$ is the carrier frequency. The distance between the ground device and satellite is a function of the zenith angle and is formulated using the cosine rule as,
\begin{equation}
    d(\varphi) = \sqrt{R_\mathrm{e}^2 + R^2 - 2R_\mathrm{e}R\cos\varphi}
\end{equation}
The Doppler shift $\nu(t)$ in \eqref{channel_Response} due to the motion of the satellite is calculated based on the change of the satellite's distance over time relative to the ground user as follows,
\begin{equation}
    \nu(t) = -\frac{f_\mathrm{c}}{c} \frac{\mathrm{d}}{\mathrm{d}t}~d(t),
\end{equation}
where $\mathrm{d}/\mathrm{d}t$ is the differentiation with respect to time, and $\mathrm{d}/\mathrm{d}t~d(t)$ is the velocity of the receiver satellite relative to the ground device. The maximum Doppler shift $\nu_\mathrm{max}$ can be analytically modeled based on the derivation of the satellite's earth-centered zenith angle relative to time multiplied by the derivation of the slant distance between the ground user and the satellite relative to the earth-centered zenith angle\footnote{Note that this model does not take into account the Earth's movement and assumes that the Earth is a perfect sphere.},
\begin{equation}~\label{DopplerMax}
    \begin{split}
    \nu_\mathrm{max} & = \pm \frac{f_\mathrm{c}}{c} \frac{\mathrm{d}\varphi}{\mathrm{d}t} \times \frac{\mathrm{d}d(\varphi)}{\mathrm{d}\varphi} \\
     & = \pm \frac{f_\mathrm{c}}{c} \omega\frac{R_\mathrm{e}R\sin\varphi_\mathrm{max}}{\sqrt{R_\mathrm{e}^2 + R^2 - 2R_\mathrm{e}R\cos\varphi_\mathrm{max}}} ,    
    \end{split}    
\end{equation}
where $f_\mathrm{c}$ is the center frequency, $c$ is the speed of light, and $\omega$ is the angular velocity of the satellite. Additionally, the received signal is assumed to be perfectly synchronized in time.

\section{Spiking-Based Detection Networks}~\label{Section_V}
The main contribution of this paper is in presenting novel spiking-based receivers in a practical S-IoT scenario. In the following section, we discuss the spiking-based neural network receivers used in this paper, namely; (i) the SNN, (ii) the CSNN, and (iii) the HybNet that was adopted from~\cite{RN648} to use spiking-based networks instead of convention DL networks.
\subsection{SNN Detector}
To implement the SNN, we utilize an open-source Python library (SnnTorch~\cite{RN674}) which encodes the input data into a spike train to accommodate spiking-based networks. The encoding is done in our work via constant current injection, where each data sample is treated as a constant current input over each time step, where 50-time steps are used. The encoding of the input can treat static data as a direct current (DC) input and consistently feed the same features to the input layer of the SNN at each time interval. However, other popular examples of encoding the input data include rate encoding, latency encoding, and delta modulation. These methods could be utilized to further extort temporal information for SNNs~\cite{RN712}. Decoding the output from spikes to real numbers is done by a process called spike decoding. Examples of decoding methods include rate decoding and latency decoding. A more expansive analysis of different encoding and decoding strategies can be found in~\cite{RN674}. For this work, mean squared error (MSE) count loss is employed, which in effect is rate decoding. MSE count loss was used as it demonstrated desirable performance when training the network. The membrane potential of the correct class is encouraged to increase the number of spikes, while the incorrect classes are encouraged to reduce the total spike count over time for each neuron to achieve high-level performance~\cite{RN693}. In addition to the SNN, a second network based on CNN called the Convolutional SNN (CSNN) is also investigated to combine the superior feature extraction and the power efficiency of the CNN relative to other DL networks, such as the ANN~\cite{RN713}, with the ultra-low power consumption of the SNN. A CSNN performs convolution operations on spikes (or events) generated by individual neurons instead of traditional continuous-valued activations. The convolutional filters in a CSNN are designed to detect specific temporal patterns in the spike sequences. This allows CSNNs to process spatiotemporal data effectively and learn to extract relevant features from input stimuli. 

To overcome the "dead neuron problem~\cite{RN674}", which significantly deteriorates performance because of the non-differentiability of spikes. The "dead neuron problem" occurs when the membrane potential is 0, thus the neuron does not fire and therefore does not contribute to the loss function in the training stage. The issue of the non-differentiability of the neuron can be addressed by smoothing out the Heaviside step function in~\ref{LIF_Eq} with the sigmoid function $\sigma(x)$, which is known as the surrogate gradient approach~\cite{RN691}. Smoothing out the Heaviside function with the sigmoid function can override the derivative of the Heaviside function. which is the Dirac-Delta function. The smoothed-out Dirac-delta function would then take time to fall to 0, therefore contributing to the training loss function. The use of surrogate gradient descent algorithms in SNNs can effectively address the dead neuron problem by allowing for gradient-based optimization and weight updates, even if some neurons are not generating spikes.


The two spiking-based networks mentioned above (SNN and CSNN) are also compared to the conventional CNN and ANN with the same dimensions as the spiking-based network counterpart. The same dimensions are used for a fair comparison between the spiking-based networks and conventional DL networks. A summary of the networks used in this work is summarized in Table~\ref{Table_CNNandSNN}. 

\begin{table*}
 \caption{ Network Summary }
 \centering
 \setlength{\tabcolsep}{4pt}
 \begin{tabular}{ccc ccc ccc ccc}
 	\hline\hline
    \multicolumn{3}{c}{ANN} & \multicolumn{3}{c}{CNN} & \multicolumn{3}{c}{SNN} & \multicolumn{3}{c}{CSNN} \\
    \cmidrule(lr){1-3} \cmidrule(lr){4-6} \cmidrule(lr){7-9} \cmidrule(lr){10-12}
    \hline
  	Layer & Shape & Activation & Layer & Shape & Activation & Layer & Shape & Neuron & Layer & Shape & Neuron \\
      \hline
            Input & $128$ & -       & Input       & $128$               & -           & Input & $128$ & LIF & Input & $128$ & LIF \\
            Dense & $800$ & tanh    & Convolution & $16$, $15 \times 1$ & ReLU        & Dense & $800$ & LIF & Convolution & $16$, $15 \times 1$ & LIF \\
            Dense & $128$ & Softmax & Max pooling & $2 \times 1$        & -           & Dense & $128$ & LIF & Max pooling & $2 \times 1$ & LIF \\
            -     & -     & -       & Convolution & $64$, $15 \times 1$ & ReLU        & -     & -     & -   & Convolution & $64$, $15 \times 1$ & LIF \\
            -     & -     & -       & Max pooling & $2 \times 1$        & -           & -     & -     & -   & Max pooling & $2 \times 1$ & LIF \\
            -     & -     & -       & Dense       & $128$               & Softmax     & -     & -     & -   & Dense & $128$ & LIF \\
 	\hline
 \end{tabular}
 \label{Table_CNNandSNN}
\end{table*}

\subsection{HybNet}~\label{Section_HybNet}
To further improve the error rate detection performance of the spiking-based receivers in noise-limited scenarios, we extend the idea from~\cite{RN648} which proposes the HybNet architecture. In this work, we adapt the HybNet architecture to the spiking-based networks so that it switches between two different detection branches; (i) a spiking-based network (in our case we select the SCNN due to its strong performance and lower power consumption compared to the ANN, CNN, and SNN, as seen in Section~\ref{energy}), and (ii) a traditional matched filter detector (in our case we utilize non-coherent detection). The advantage of utilizing the HybNet architecture is that in noise-limited scenarios the matched filter path is chosen by the supervisor switching network to detect the signal. Conversely, when the receiving signals are interference-limited, detection using the spiking-based network is chosen by the supervisor switching network. Note that this supervisor switching network is trained with an SCNN network, which we dub as the \textit{Selector SCNN}, and the network structure is listed in Table~\ref{Table_selector}. 

\begin{figure}[!t]
{\centering
\includegraphics[width=\linewidth]{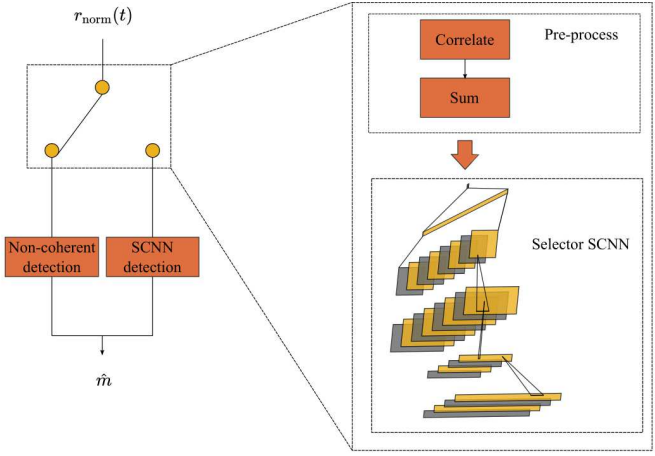}
\caption{ An illustration of the HybNet architecture used in this paper. }
\label{Fig_Models}}
\footnotesize
\end{figure}

In order to train the selector network, each target chirp-based symbol is cropped and labeled either as; (i) \emph{Minimal interference} or (ii) \emph{Interference-limited}. When the target symbol power is higher than the interference signal power (i.e., SIR $>$ 0 dB), the symbol is labeled as ``Minimal interference''. Alternatively, if the power of the interfering transmissions is larger than the power of the target signal, the symbol is labeled as ``Interference-limited''. When this signal is passed into the \textit{Selector SCNN}, the signal is passed to be detected by the non-coherent detector (outlined in subsection~\ref{nCoh_subsection}) if it is classified as ``Minimal interference'' and if it is classified as ``Interference-limited'' then the symbol is detected by the SCNN in this work. An illustration of the utilized switching architecture is shown in~Fig.~\ref{Fig_Models}. 

\begin{table}
 \caption{Selector SCNN}
 \centering
 \begin{tabular}{ccc}
 	\hline\hline
 	Layer & Shape & Neuron \\
 	\hline
 	Input & $128$ & LIF \\
          Convolution & $4, 15 \times 1$ & LIF \\
          Max pooling & $2 \times 1$ & LIF \\
          Convolution & $8, 15 \times 1$ & LIF \\
          Max pooling & $2 \times 1$ & LIF \\
          Dense & $2$ & LIF \\
 	\hline
 \end{tabular}
 \label{Table_selector}
\end{table}  

\subsection{Dataset Creation and Training}
\begin{figure}[!t]
{\centering
\includegraphics[width=\linewidth]{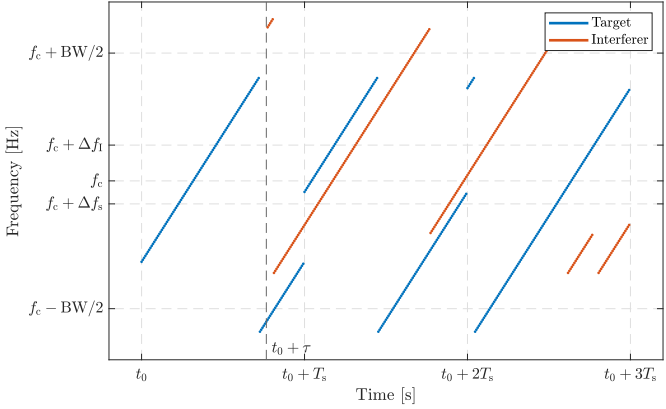}
\caption{ A snapshot of a three-symbol wide spectrogram of an example chirp-based signal with interference for another chirp-based signal. The interference signal is time-shifted by $\tau$, has a Doppler shift of $\Delta_\mathrm{I}$, and the target signal has a Doppler shift of $\Delta_\mathrm{s}$.}
\label{Fig_LoRa}}
\footnotesize
\end{figure}

To ensure the desired performance of the DL/spiking-based receivers, the networks need to be trained on a dataset that contains emulated signals that exhibit the behavior of the received chirp-based IoT uplink signals in a practical LEO satellite scenario. For creating the dataset for training the DL/spiking-based receiver networks for signal detection, chirp I/Q signals are emulated, where the signals in the dataset are generated in our satellite scenario emulator. Each satellite is designated an empty vector 1D with a length determined by the time-step $dt$ of the simulation. Next, the satellite's empty vector is populated with signals from the active ground user devices within its coverage area. Each signal contains an n-length random symbol sequence ${M = \{m_\mathrm{1},m_\mathrm{2},...,m_\mathrm{n}\}}$ that has a controlled transmit power $p_\mathrm{Tx}$ and has a random time offset $\tau$. Every transmitted signal is impaired by the emulated ground-to-satellite channel effects which are described in~\ref{Section_II}. Additionally, a randomly chosen target signal is chosen as the target signal and is appended with a preamble for synchronization using the developed method detailed in Section~\ref{Section_VII}. After synchronization, the target signal is normalized by dividing by the target signal power $p_\mathrm{s}$ which is estimated from the preamble. The normalization is shown mathematically as follows,
\begin{equation}
    r_\mathrm{norm}(t)=\frac{r(t)}{\sqrt{p_\mathrm{s}}} .
\end{equation}
The reason for normalizing the signal power to $p_\mathrm{s}=1$ is to maintain a uniform magnitude of the target signal power during the training of the receiver. This is because scaling the input has been proven to improve performance~\cite{RN648, norm}. Furthermore, normalization provides the receiver network with information about which signal is the target signal and which is the interfering signal. Finally, each symbol is cropped out and is represented as follows,
\begin{equation}
    Y_\mathrm{k} = |y(t)S^*_k(t)| ,
\end{equation}
where $y(t)$ denotes the received MFSK chirp-based signal and $S_k(t)$ is the MFSK reference signal with frequency shift $k\Delta_{\mathrm{f}}$, this is the same equation as in Eq.~\ref{lora_corr} without the argmax function. Each training symbol is then labeled and can be represented as follows,
\begin{equation}
    T = \{(Y_\mathrm{1},m_\mathrm{1}),(Y_\mathrm{2},m_\mathrm{2}),...,(Y_\mathrm{k},m_\mathrm{k})\} ,
\end{equation}
where $m$ is the symbol label. An illustration of an example chirp signal is shown in Fig.~\ref{Fig_LoRa}.

To train the \textit{Selector SCNN} the same dataset that was used to train the DL/spiking-based receivers is utilized. However, only two labels were used, \emph{Minimal interference} or \emph{Interference-limited}. The labeling procedure is described in Section~\ref{Section_HybNet}.

\section{Chirp-Based Signal Emulation}~\label{Section_VI}
For this work, we incorporate the LoRa modulation technique as an example of chirp-based modulation due to its widespread use for LPWAN applications. We utilize the LoRa MATLAB emulator~\cite{Emulator} developed earlier by our team which consists of two components: (i) signal generator, and (ii) conventional detector based on non-coherent detection. Note, for this work, there is no additional error correction involved, just the generation of a chirp-based packet with a random symbol payload, and a receiver that detects the payload packet with a non-coherent receiver compared with the developed DL/spiking-based receiver. Signal detection performance with error correction would decrease the error rate. As such, in this paper, we show signal detection performance without error correction as the lower bound. 

\subsection{LoRa Modulation}
LoRa uses the chirp spread spectrum (CSS) technique~\cite{7797659} to modulate the transmitted symbols. This is achieved by shifting the start of the chirp according to the transmitted symbol value, and then the frequency is cyclically swept within a given bandwidth B. According to LoRaWAN specifications only a few discrete values of the bandwidth are permitted; where $B \in \{125,250,500\}$~kHz. Another parameter that controls the chirp rate of each symbol is called the Spreading Factor (SF), which is also restricted in the LoRaWAN specifications to $\text{SF} \in \{7,8,9,10,11,12\}$. The chirp rate can be calculated with the SF as $\frac{d\phi}{dt} = \frac{SF}{T_\mathrm{chirp}}$, where $T_\mathrm{chirp}$ is the time of the chirp. In addition, the SF defines the number of possible values that can be encoded by the symbol, which is given by $\mathrm{M} = 2^\mathrm{SF}$. 

\subsection{LoRa Detection}~\label{nCoh_subsection}
The detection of a LoRa signal involves a two-step process; in the first step, the signal is dechriped by mixing the LoRa signal with an inverted chirp. The dechirping process converts the LoRa signal into a multiple frequency-shift keying (MFSK) signal, which is a modulation type that encodes symbols with M equally spaced frequency tones. The second step can then be detected either using conventional coherent or non-coherent detection methods. The coherent detection method based on matched filtering is proven to be an optimal detector under AWGN channel conditions~\cite{simon2005digital} and is also shown to  achieve better performance compared to the non-coherent detector in the presence of LoRa-on-LoRa interference~\cite{unknown}. However, coherent detection requires perfect synchronization in both time and frequency~\cite{simon2005digital} for coherent detection to perform well~\cite{RN525}. Therefore, coherent detection is not appropriate for a LEO satellite communication scenario due to the high Doppler frequency shift on the transmitted signal. Consequently, we use non-coherent detection in this paper.  

To describe non-coherent detection, the square-law (envelope) matched filter detector can be used~\cite{simon2005digital}. The signal is mixed with (i.e., mathematically multiplied by) by the conjugate of all possible realizations of the chirp-based signal and the maximum magnitude denotes the symbol estimate $\hat{m}_{\text{ncoh}}$ as follows,
\begin{equation}
    \hat{m}_{\text{ncoh}} = \argmax_k |y(t)S^*_k(t)|,
\end{equation}~\label{lora_corr}
where $y(t)$ denotes the received MFSK chirp-based signal and $S_k(t)$ is the MFSK reference signal with frequency shift $k\Delta_{\mathrm{f}}$, represented as $z_{\mathrm{k}}(t) = \exp\left(j2\pi k\Delta_{\mathrm{f}} t\right)$ and $k$ is an integer representing all symbol possibilities ${k = \{0,1, ... , M-1\}}$. $\Delta_\mathrm{f}$ is the frequency step between the MFSK shifts where the frequency step for chirp-based symbols is equal to the symbol rate $B/M$. 

\section{Frame Synchronization}~\label{Section_VII}
\begin{figure}[!t]
{\centering
\includegraphics[width=\linewidth]{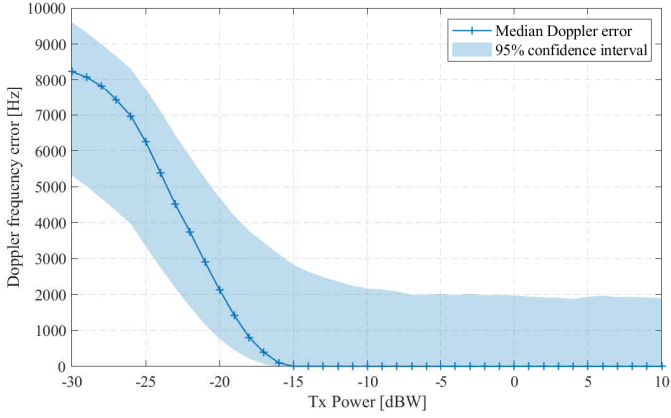}
\caption{A plot of the median frequency synchronization error against different transmit powers. The spatial duty cycle $D$ was chosen as $1\times 10^{-6}$.}
\label{Fig_Doppler}}
\footnotesize
\end{figure}

The preamble in the IoT frame is used to train the local chirp generator on finding the start of the chirp. The preamble contains no payload data and is just a sequence of symbols that are appended at the beginning of the frame. Given the large Doppler shifts in LEO orbits, we propose a synchronization method based on time-frequency matched filtering (inspired by range Doppler matched filtering in radar signal processing~\cite{Cook1967RadarSA}) by matching the ideal templates of the known preamble each having a slightly different Doppler shift. Mathematically the time domain signal of preamble templates with different Doppler shifts can be expressed as follows,
\begin{equation}
    K = x_\mathrm{pre}\exp(j2\pi \kappa(t)) ,
\end{equation}
where $x_\mathrm{pre}$ is the ideal preamble template and $\kappa(t)$ is the vector of different Doppler shifts. We first perform a coarse search with a Doppler frequency shift spacing of 100 Hz, which is denoted as $\eta_\mathrm{c}$. Then when the best match is found, a fine search is then performed around the best candidate with a frequency spacing of 1 Hz, which is denoted as $\eta_\mathrm{f}$. We make $K \in M \times N$ matrix, where $N$ is the number of temporal samples in the signal, and $M$ is the resolution of the search space. The resolution for the coarse search is equal to the length of $\kappa(t)$, which can be obtained as,
\begin{equation}
    M_\mathrm{c} = \kappa_\mathrm{len} = 2\frac{\nu_\mathrm{max}}{\eta_\mathrm{c}} ,
\end{equation}
where the $\nu_\mathrm{max}$ is the maximum Doppler shift (calculated with Eq.~\ref{DopplerMax}). The resolution of the fine search is taken as,
\begin{equation}
M_\mathrm{f} = 2\eta_\mathrm{c} .
\end{equation}
Then, to create a 2D time-frequency plot of the similarity values, we employ a 2D cross-correlation as follows,
\begin{equation}
    C(k,l) = \sum_{m=0}^{M-1}\sum_{n=0}^{N-1}K(m,n)y^*(m-k,n-l) ,
\end{equation}
where $y^*(.)$ corresponds to the complex conjugate of the received signal vector, and $y$ is a $P \times Q$ matrix, where $P$ is equal to $M_\mathrm{c}$ when performing coarse search and $P$ is equal to $M_\mathrm{f}$ when performing the fine search. $Q$ is equal to the number of temporal samples in $x_\mathrm{pre}$. Matched filtering in this way would yield a 2D array of similarity values where the maximum value is taken to be the best match. Finding the best match after the fine search would give an accurate time and frequency synchronization with the target IoT packet. The performance of the frame synchronization method is shown in Fig.~\ref{Fig_Doppler} where the synchronization accuracy decreases as the transmit power decreases due to the lower SNR.

\begin{table}
 \caption{Simulation Parameters}
 \centering
 \begin{tabular}{l c c}
 	\hline\hline
 	Parameter & Symbol &Value \\
 	\hline
 	Satellite constellation & - & Walker Delta \\
 	Satellite orbital inclination & $i$ & 53$\degree$ \\
 	Satellite orbital planes & $P$ & 16 \\
 	Satellite phasing parameter & $F$ & 8 \\ 
 	Satellite constellation altitude & $h$ & 1200~km \\
 	Number of satellites & $N_\mathrm{s}$ & 384 \\
        Time-step & $dt$ & 0.1275~s \\
 	Density of IoT devices & $\lambda_\mathrm{o}$ & $1~\text{units~per~km}^2$ \\ 
        Duty cycle & $D$ & $1 \times 10^{-6}$ \\
 	Earth’s radius & $R_\mathrm{e}$ & 6371~km \\
 	Center frequency & $f$ & 915~MHz \\
 	Transmit power & $P_\mathrm{t}$ & 0~dBW \\
    Satellite antenna beamwidth & $\psi$ & 80$\degree$ \\
    Satellite antenna Gain & $G$ & 9~dBi \\
    Average noise power & $W$ & -117~dBm \\
    LoS probability parameter & $\zeta$ & 0.4 \\
    LoS excess mean & $\mu_\mathrm{LoS}$ & 0 \\
    LoS excess standard deviation & $\sigma_\mathrm{LoS}$ & 1 \\
    NLoS mean & $\mu_\mathrm{NLoS}$ & 8 \\
    nLoS standard deviation & $\sigma_\mathrm{NLoS}$ & 10 \\
    Spreading factor & SF &  7  \\
    Bandwidth & $B$  & 125~kHz \\
    Initially sampling rate & $f_\mathrm{s}'$ & 1~MHz \\
    Sampling rate & $f_\mathrm{s}$ & 250~kHz \\
    Symbol time & $T_s$ & 1.024~ms \\
    Symbols per packet & $N_\mathrm{p}$ & 10 \\
    Symbol index & $m$ & -\\
    Number of training frames &-& 160,000 \\
    Number of test frames &-& 40,000 \\
 	\hline
 \end{tabular}
 \label{Table_Sim}
\end{table}

\section{Results and Discussion}~\label{Section_VIII}
In this section, we provide Monte Carlo simulation results of the emulated satellite system, where the simulation parameters are listed in Table~\ref{Table_Sim}. Chirp-based signals are emulated and passed through the simulated ground-to-satellite channel, where the spatial duty cycle $D$ dictates the number of transmissions sent each second. The spatial duty cycle $D$ was arbitrarily chosen as $1\times 10^{-6}$ it allows for a balance between interference-limited and noise-limited scenarios experienced by the satellite receivers. Bear in mind that to create accurate signals, an oversampled signal vector is constructed by sampling at $f'_\mathrm{s} = 1$~MHz. Each transmitted chirp-based signal is impaired by the emulated ground-to-satellite channel. After all the signals from the active devices within a satellite footprint at a certain time period are superimposed onto the signal vector. The vector is downsampled to twice the Nyquist sampling frequency of $f_\mathrm{s} = 250$~kHz to increase robustness against noise and Doppler. Note that for this work, we emulate LoRa signals with a $\text{SF} = 7$, and a $B = 125$~kHz. A higher SF can be trained on, however, this would increase the training dataset size so that for every increment of SF, the size of the training dataset be cumulatively doubled. This increase in dataset size is needed to support the additional numbers of symbol values a symbol can encode, which is equal to $2^\mathrm{SF}$. Furthermore, a more complex DL network would also be required. The CNN and ANN are trained using stochastic gradient descent with momentum over $50$ epochs with a learning rate of $1e^{-3}$ and a mini-batch size of $256$. On the other hand, the SNN, CSNN, and Selector CSNN are trained with a batch size of 128 is used, with the Adam optimizer and a learning rate of $5e^{-4}$ over $50$ epochs. These hyperparameters for training were chosen after performing manual optimization and the parameters showed desirable performance of the networks. 

\begin{figure}[!t]
    {\centering
        \includegraphics[width=\linewidth]{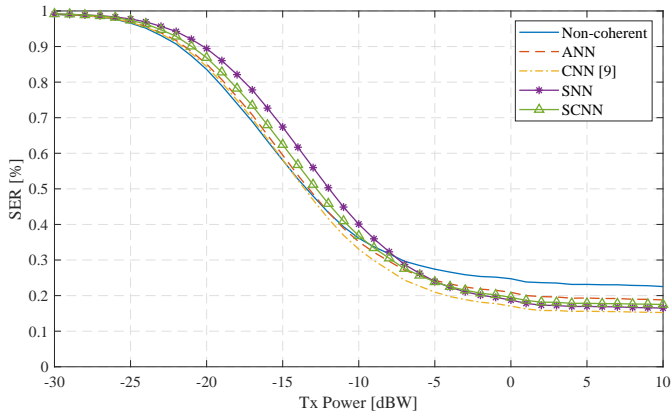}
        \caption{ A plot of the average SER against different transmit powers, where plots are shown with a spatial duty cycle $D$ of 1e-6$\%$. Manual Doppler compensation and synchronization are also performed using the preamble. }
        \label{Fig_SERvPtx_D_wpre}}
    \footnotesize
\end{figure}

To demonstrate the effectiveness of the DL and spiking network for signal detection in a practical scenario, the SER performance is shown in Fig.~\ref{Fig_SERvPtx_D_wpre}. Synchronization in time and Doppler compensation is performed using the time-frequency matched-filter synchronization method discussed in Section~\ref{Section_VII}. Additionally, the signal is normalized to the average power of the preamble. The SER curve decreases as the transmit power increases, for both non-coherent as well as for the DL/spiking-based receivers as the system moves from noise-limited to interference limited. However, the DL/spiking-based methods show a greater performance relative to the non-coherent receiver at the region of the SER plateaus, and the system becomes interference-limited.

\begin{figure}[!t]
	{\centering
		\includegraphics[width=\linewidth]{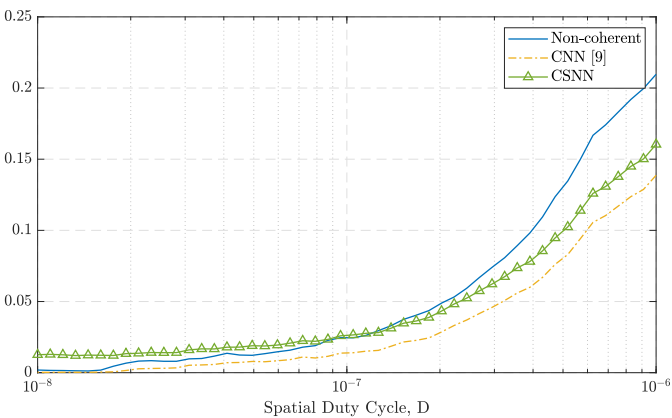}
		\caption{ A plot of the average SER against different spatial duty cycles $D$, where the transmit power is fixed at 0 dBW. Manual Doppler compensation and synchronization are also performed using the preamble. }
		\label{Fig_SERvsDC}}
	\footnotesize
\end{figure}

The SER performance is also recorded as the spatial duty cycle $D$ increases and is shown in Fig.~\ref{Fig_SERvsDC}. From this plot is evident that performance improvement between conventional detection and DL/spiking-based detection increases as the spatial duty cycle increases. The improvement comes from the ability of DL and spiking-based networks to correctly detect symbols in the presence of co-channel interference. 

From the SER performance plots, we can see that the CNN has the highest performance compared to the other networks discussed in this work. However, the spiking-based networks still outperform conventional detection and have ultra-low power consumption which is further analyzed in the subsection~\ref{energy}.

\subsection{HybNet Performance}
\begin{figure}[!t]
{\centering
\includegraphics[width=\linewidth]{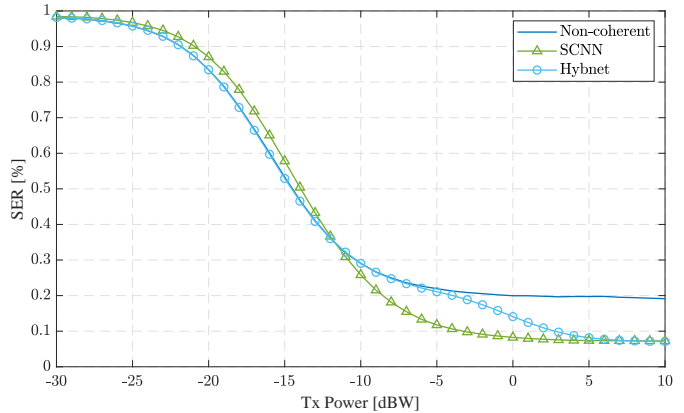}
\caption{ A plot of the performance of the HybNet framework, where plots are shown with a spatial duty cycle $D$ of 7e-7$\%$ without Doppler. }
\label{Fig_Hyb}}
\footnotesize
\end{figure}

The performance of the \emph{HybNet} architecture is shown in Fig.~\ref{Fig_Hyb}. From the plot, HybNet switches between being purely non-coherent detection in noise-limited scenarios and being detected with the SCNN in interference scenarios. The switching then allows for an average decrease in the SER relative to both the SCNN and noncoherent detection. However, at the transition between noise-limited and interference limited the Selector SCNN fails to correctly classify which detection pathway is more efficient. 

\subsection{Energy Consumption}~\label{energy}
To demonstrate the advantages of spiking-based learning against conventional learning in terms of power efficiency, benchmarks for energy per detection are recorded. The energy is recorded using the Keras-Spiking Python package from Nengo~\cite{KerasSpiking}. The following assumptions are made when estimating the energy used by the proposed model on a particular device:
The Energy consumption approximation is calculated based on the energy used per operation, where the sources for the GPU~\cite{RN705} and for Loihi~\cite{Loihi} are used. Overhead is also not considered such as the energy required to transfer the data to be classified by the network. Only the energy consumption of the components in each network is considered. 

From the Table.~\ref{Table_Energy}, The spiking-based networks use a few orders of magnitude less power per detection compared to their conventional DL network counterparts. These results reinforce the idea that a spiking network could be used for signal detection, particularly in resource-constrained applications such as in a LEO satellite.

\begin{table}
 \caption{Energy Consumption}
 \centering
 \begin{tabular}{l ccc}
 	\hline\hline
    \multicolumn{1}{l}{} & \multicolumn{3}{c}{Energy per detection (J)} \\
    \cmidrule(lr){2-4}
    \hline
 	Network & CPU & GPU & Loihi  \\
 	\hline
 	ANN & 1.14$\times10^{-1}$ & 3.99$\times10^{-3}$ & - \\
 	CNN & 7.41$\times10^{-3}$ & 2.59$\times10^{-4}$ & - \\ 
 	SNN & - & - & 4.15$\times10^{-4}$ \\
 	CSNN & - & - & 9.33$\times10^{-6}$ \\
        Selector CSNN & - & - & 1.23$\times10^{-6}$ \\
 	\hline
 \end{tabular}
 \label{Table_Energy}
\end{table}

\section{Conclusion}~\label{Section_IX}
In this paper, we presented a spiking network-based receiver for detecting IoT signals in a satellite IoT uplink scenario. For the receiver, we investigated the spiking neural network (SNN) and the convolutional SNN (CSNN) and compared the error rate performance against conventional ANN and CNN receivers, as well as conventional non-coherent detection. The findings reveal that both spiking-based and DL receivers exhibit resilience to co-channel interference. Notably, the spiking-based networks display impressive detection performance in high interference scenarios, while consuming several orders of magnitude less power per detection than traditional ANN and CNN receivers. To further improve the detection performance of the spiking networks, we adopt the HybNet framework from our previous work to switch between the traditional detection methods and the spiking-based receiver. 

\bibliographystyle{ieeetr}
\bibliography{main}

\clearpage

\end{document}